\documentclass[prb,preprint]{revtex4-1} 

\usepackage{amsmath}  
\usepackage{amsfonts} 
\usepackage{graphicx} 
\usepackage{caption}
\usepackage{subcaption}
\usepackage{enumitem}
\usepackage{framed}
\usepackage{xcolor}

\begin{document}


\title{A Complete Analytic Gravitational Wave Model for Undergraduates}
\author{Dillon Buskirk \footnote{Undergraduate Student, email: {\href{mailto:buskirk16@live.marshall.edu}{buskirk16@live.marshall.edu}}}}
\author{Maria C.~Babiuc Hamilton \footnote{email: \href{mailto:babiuc@marshall.edu}{babiuc@marshall.edu}}}
\affiliation{Department of Physics, Marshall University, Huntington, WV 25755, USA}

\begin{abstract}
Gravitational waves are produced by orbiting massive binary objects, such as black holes and neutron stars, and propagate as ripples in the very fabric of spacetime. 
As the waves carry off orbital energy, the two bodies spiral into each other and eventually merge. They are described by Einstein's equations of General Relativity. 
For the early phase of the orbit, called the inspiral, Einstein equations can be linearized and solved through analytical approximations, while for the late phase, near the merger, we need to solve the fully nonlinear Einstein's equations on supercomputers. 
In order to recover the gravitational wave for the entire evolution of the binary, a match is required between the inspiral and the merger waveforms. 
Our objectives are to establish a streamlined matching method, that will allow an analytical calculation of the complete gravitational waveform, while developing a gravitational wave modeling tutorial for undergraduate physics students. 
We use post-Newtonian (PN) theory for the inspiral phase, which offers an excellent training ground for students, and rely on \texttt{Mathematica} for our calculations, a tool easily accessible to undergraduates. For the merger phase we bypass Einstein's equations by using a simple analytic toy model named the Implicit Rotating Source (IRS). 
After building the inspiral and merger waveforms, we construct our matching method and validate it by comparing our results with the waveforms for the first detection, GW150914, available as open-source. 
Several future projects can be developed based from this project: building complete waveforms for all the detected signals, extending the post-Newtonian model to take into account non-zero eccentricity, employing and testing a more realistic analytic model for the merger, building a separate model for the ringdown, and optimizing the matching technique.
\end{abstract}
\maketitle 

\section{Introduction}
\label{sec:Intro} 
Why are gravitational waves important? They ripple through the very fabric of space-time and, like other kind of waves, carry information about their sources: cataclysms of cosmic proportion, such as colliding black holes, exploding supernovae, and even the origins of the universe. 
This information, once decoded, will enable us to answer deep and fundamental questions about the Universe and the nature of space and time.

The first direct detection of gravitational waves happened on September 14,  2015, thanks to the  precise instruments of the Laser Interferometer Gravitational-wave Observatory (LIGO).
This discovery, known as the GW150914 event~\cite{abbott1602}, came from the collision of two black holes. 
The announcement came at the beginning of 2016, almost as if to celebrate the $100^{th}$ anniversary of Einstein's Theory of General Relativity. 
Three scientists who played an instrumental role in this discovery received the 2017 Nobel prize in physics for their contribution to one of the most important achievements in the history of science.

Several more detections followed suit: GW151226~\cite{abbott1606}, GW170104~\cite{abbott1706}, GW170608~\cite{abbott1708}, GW170814~\cite{abbott1709}, and GW170817~\cite{abbott1710}.
The last event is also known as the \emph{golden binary}, because the collision gave off -- besides gravitational waves -- electromagnetic radiation across the spectrum, and hundreds of Earth masses of precious and heavy elements.

With these recent discoveries, the era of gravitational and multi-messenger wave astronomy has begun, and  
crucial to the success of this new science is the development of a reliable pipeline of well-prepared and capable researchers, ready to move this field forward.  

The Einstein field equations of general relativity are essential for the correct modeling of gravitational waves.  
These complicated partial differential equations are extremely challenging to solve analytically.
For example, in order to obtain the correct gravitational wave signal from two orbiting stars, the solution would have to contain everything, starting with the birth of the binary from the interstellar gas, continuing with the evolution to the merger, and finishing with its collapse into a black hole, with an emission of gravitational radiation~\cite{poissongravity}.
So far, no exact solution exists for such complex systems. 
In fact, only a few exact solutions of Einstein's equations are known~\cite{wald84}. 
Therefore, in order to calculate the gravitational waves, we need to resort either to numerical simulations~\cite{pretorius05} or to analytical approximations~\cite{blanchet13}. 

This research started as senior undergraduate capstone project during the academic year of 2017-2018, when the analytical models were implemented, and applied to the first 5 gravitational wave detections, without being tested. 
During the summer of 2018, we continued with work on the matching technique, in order to obtain complete analytical gravitational waveforms for the entire evolution of the binary, and validate it with the template for the GW150914.
We will give here an extended report of this work and the results obtained.

We start by presenting in Sec.~\ref{sec:Models}, the two analytical approaches we are using to calculate the gravitational waves: the \emph{post-Newtonian} (PN), and the \emph{Implicit Rotating Source} (IRS) models, explaining their domains of applicability. 
In Sec.~\ref{sec:Implementation}, we elaborate the algorithm employed in numerically determining the evolution variables that enter in the calculation of the gravitational wave amplitude, using \texttt{Mathematica}~\cite{MATH}, and we calculate the strain for a fiducial waveform, for both the inspiral and the merger.
In Sec.~\ref{sec:Matching}, we concentrate our efforts in finding a simple and effective method for matching the end of the inspiral with the beginning of the merger, and we build a complete analytic waveform for the whole evolution of a fiducial binary configuration.  
Furthermore, we investigate whether the model is correct by calculating the gravitational waveform for the binary configuration corresponding to the GW150914 event, and testing it against the open-source template for the GW150914 strain~\cite{LOSC}. We obtain a very good overlap between our calculated strain and the template, which proves that our calculations are correct.
Finally, in Sec.~\ref{sec:Conclusions}, we summarize our work and give a short outline of future undergraduate projects that can be developed. 

Throughout this work we are using geometric units, such that $G=c=1$.
Therefore, the equation for the Newtonian gravitational field simplifies to: $\Phi = -M/r$.
Those units are commonly used in general relativity calculations and undergraduates are likely unfamiliar with them. 
It is useful to review the relationship between those units and the \emph{International System of Units} (SI), because it it is not intuitive and its usefulness is not easy to grasp.
We describe in Appendix~\ref{appx:Units} how to convert between SI and geometric units, giving examples on how this allows us to measure mass in seconds, distance in mass, etc. 
Among the symbols used in the paper are the following: the total mass of the binary $M=m_1 + m_2$, with $m_1$ and $m_2$ as the individual masses, the symmetric mass ratio $\eta=m_1m_2/M^2$, the orbital separation between the location of the centers of the stars in the binary $r$, the overall orbital velocity $v$, the orbital angular velocity $\omega$, the frequency of the gravitational waves $f_{GW}$, the phase $\Phi$, the distance from the detector to the binary $R$, the polarization modes of the strain $(h_{\times},h_+)$, and the amplitude of the wave $A$. Appendix~\ref{appx:PNcoeffs} and Appendix~\ref{appx:Mcoeffs} contain the PN and gIRS coefficients employed in our calculations. 
In Appendix~\ref{appx:Procedures} we present a short step by step tutorial that if followed, will enable 
the reader to change the necessary parameters in order to obtain complete gravitational waveforms for different binary configurations.  
\section{The Analytical Models}
\label{sec:Models}
Analytical models of compact binary inspiral and merger are commonly used in combination with numerical simulations, in order successfully to build a large bank of accurate waveform templates ~\cite{abbott1602, abbott1606, abbott1706, abbott1708, abbott1709, abbott1710, east1212, huerta1609, blackman1705}. 
The best-known analytical procedure of calculating gravitational waves uses the PN theory, which was originally developed by Einstein in order to find solutions to his field equations of general relativity.  
This theory shows how to construct perturbative solutions of Einstein's equations as a series of successive approximations in powers of $v/c$, in the case of slow motion, large separation, and weak gravitational fields~\cite{einstein37}. 
The ratio of the velocity of the source to the speed of light is called the post-Newtonian parameter, $x=v^2/c^2$.  
Although its validity is limited to weak fields, the post-Newtonian theory offers remarkably accurate predictions for the gravitational radiation emitted by compact binary systems~\cite{blanchet13, clifford}. 
We will show below how to use the post-Newtonian approach to calculate the gravitational waves emitted during the inspiral of two black holes.
We must keep in mind that this approach breaks down for large speeds, as $x$ gets close to unity, and it cannot be applied at the merger, where we have to rely either on numerical relativity or on analytical toy models. 
One of the simpler and well known analytical ansatz used to model the merger case is the  \emph{Implicit Rotating Source} model, as described in~\cite{buonanno0610, east1212, huerta1408, huerta1609}. 
We will use the \emph{generic} IRS toy-model for the merger~\cite{huerta1609, baker2008}, which is tuned to numerical relativity, it is easy to use, and gives satisfactory results.
Other merger models are presented in~\cite{blackman1705}. 

Once each phase is modeled, a match is required between the post-Newtonian waveform describing the inspiral and the merger waveform obtained with the gIRS toy-model. 
We devise a simple and efficient stratagem for the matching region between the PN generated waveform and the gIRS wave model, very close to the merger, and we prove its validity. 
\subsection{The Inspiral Model}
We know that in Newton's law of universal gravitation, a binary system is stable, and will orbit indefinitely with constant frequency, without emitting gravitational waves, or shrinking their orbit.
However, Einstein's general theory of relativity predicts that the two orbiting stars will gradually lose orbital energy through emission of gravitational waves and will come closer together until they merge to form one single star of larger mass, or a black hole. 
In this situation, the energy will decrease in time, and the equation for the energy balance describing this behavior will be written as:
\begin{equation}
  \label{eq:eBalance}
  \frac{dE}{dt} = -\mathcal{F},
\end{equation}
where $E$ is the energy of the binary and $\mathcal{F}$ is the flux of the emitted gravitational waves. 
Even the Earth--Sun system emits gravitational waves, but their effect is too weak ($10^{-24}$ loss relative to the total orbital energy) to be taken into account.

In the post-Newtonian approximation, the equations of general relativity take the form of the familiar Newtonian two--body equations of motion, in the limit $v/c\rightarrow0$, called the \emph{weak field} limit.  
A correction of order $(v/c)^n$ to the Newtonian equation of motion is counted as an $n/2$ order in the PN expansion. For example, the two-body equation of motion becomes:
\begin{equation}
\frac{d v}{dt} = -G \frac{M}{r^2}[1 +\frac{1PN}{c^2}  + \frac{1.5PN}{c^3} + \frac{2PN}{c^4} + \frac{2.5PN}{c^5} + ...]
\end{equation}
At each post-Newtonian expansion  we unravel new physics beyond the Newtonian realm. 
For example, the $1^{st}$ order recovers orbit precession, the $1.5^{th}$ order describes spin-orbit interaction, and the $2^{nd}$ order spin-spin coupling dynamics. 
The  orbital decay with emission of gravitational waves appears from the $2.5^{th}$ order onward. 
The current state of the art in the post--Newtonian expansion is $3.5^{th}$ order~\cite{blanchet13, futamase}.

Using Kepler's third law of planetary motion $(\omega^2 r^3 = G{M})$ and writing the orbital velocity as $v = \omega r = (G {M} \omega)^{1/3}$, we obtain the following important relationship between the post-Newtonian parameter $x$ and the orbital angular velocity $\omega$: 
\begin{equation}
 x = \frac{v^2}{c^2}=\frac{(G{ M}\omega)^{2/3}}{c^2}
\label{eq:xom}
\end{equation}
In geometrical units, eq.~\eqref{eq:xom} becomes $x=(M\omega)^{2/3}$. 
Using the chain rule, we rewrite the energy balance eq.~\eqref{eq:eBalance} in terms of $x$ as:
\begin{equation}
  \label{eq:OmEv}
  \frac{dx}{dt} = -\frac{\mathcal{F}}{dE/dx}.
\end{equation}
There are several well known ways of solving eq.~\eqref{eq:OmEv}, referred to as the Taylor T1 through T5 approximants~\cite{ajith2011ec}.
We will use the Taylor T4 approximant method, which was reported to give better agreement with numerical relativity than other approximants~\cite{blackman1705, hinder0806} for binaries with comparable mass. 
This method expands~\eqref{eq:OmEv} in post-Newtonian powers of $x$:
\begin{equation}
\frac{dx}{dt} =  \frac{dx^{0\,\rm{PN}}}{dt} x^5 + \frac{dx^{1\,\rm{PN}}}{dt} x^6 + \frac{dx^{2\,\rm{PN}}}{dt} x^7 + \frac{dx^{3\,\rm{PN}}}{dt} x^8 + \frac{dx^{\rm HT}}{dt}.
\label{eq:xdot}
\end{equation}
Here $\dot{x}_{\rm HT}$ stands for \emph{hereditary} terms, and represent the higher order post-Newtonian non-linear terms, which depend on the dynamics of the system in its entire past, the so called \emph{tails} and \emph{tails-of-tails} terms that account for the nonlinear interaction between the gravitational waves and the spacetime itself.
Each post-Newtonian term in eq.~\eqref{eq:xdot} is further expressed as a power series in $x$, truncated at the appropriate order.
We will use up to the $6PN$-order terms, which are the highest order calculated for a quasi-circular orbit~\cite{huerta1609}. 
The Taylor-T4 approximant in the quasi-circular limit has the following expression:
\begin{equation}
\label{eq:pre_hyb}
M\frac{\mathrm{d}x}{\mathrm{d}t}\bigg|^{\rm 6PN} 
= \tfrac{64}{5} \eta x^5 \left (1 + \sum_{k=2}^{12} a_{\tfrac{k}{2}}x^{\tfrac{k}{2}} \right).
\end{equation}
This equation, when integrated, gives the evolution of the post--Newtonian variable $x$. 
We give in Appendix~\ref{appx:PNcoeffs} the formulas for the expansion coefficients $a_i$, and make available the \texttt{Mathematica} script where we implemented those coefficients and integrated eq.~\eqref{eq:pre_hyb} numerically.
Once $x$ is known, we will obtain the orbital phase by integrating the equation:
\begin{equation}
  \label{eq:PhiEv}
  M\frac{d\Phi_{orb}}{dt} =M\omega_{orb}=x^{3/2}.
\end{equation}
The binary orbit shrinks during the evolution, therefore the separation depends on time as well.
In order to describe this, we can calculate $r$ directly from the post-Newtonian parameter $x=\frac{v^2}{c^2}=\frac{\omega^2r^2}{c^2}$, using Kepler's third law $\omega^2 r^3 = G{M}$, to obtain $r(t) = M x(t)^{-1}$ in geometrical units.
We will push the precision in the calculation of the separation even further, by applying post-Newtonian corrections up $3 PN$~\cite{hinder0806}:
\begin{equation}
\label{eq:r}
r = M(r^{0\,\rm{PN}}x^{-1} + r^{1\,\rm{PN}} + r^{2\,\rm{PN}}x+r^{3\,\rm{PN}}x^2),
\end{equation}
with the terms $r^{i\,\rm{PN}}$ included in Appendix~\ref{appx:PNcoeffs}. 

Once the evolution of the orbital phase and the separation are known, we can construct the amplitude of the gravitational wave as a combination of two independent states of polarizations, similar to electromagnetic waves. 
For this, we use the general formula: 
\begin{eqnarray}
h_{+}& =&-\frac{M\eta}{R}
\bigg\{(\cos ^2 \theta +1 )
\bigg[ 
\left(-\dot r^2+r^2\dot \Phi^2+\frac{M}{r}\right) \cos 2 \Phi 
+ 2 r \dot r \dot \Phi \sin 2\Phi 
\bigg ] 
+ \left(-\dot r^2+r^2\dot \Phi^2+\frac{M}{r}\right) \sin ^2 \theta 
\bigg\} 
\nonumber \\
h_{\times} &=&-2\frac{M\eta}{R}\cos \theta
\bigg[\left(-\dot r^2+r^2\dot \Phi^2+\frac{M}{r}\right) \sin 2\Phi 
- 2 r \dot r \dot \Phi \cos 2\Phi \bigg].
\label{eq:h_gen}
\end{eqnarray}
The two polarization modes are denoted $h_{+}$ and $h_{\times}$, to emphasize that the angle between them is $\pi/4$, and not $\pi/2$, like with electromagnetic waves. 
The amplitude of the gravitational waves depends on the orientation of the binary with respect to the detector. 
When working in the detector's coordinate system, also called the \emph{fundamental frame},  we have to take into account the {\em inclination angle} $\theta$, which is the angle between the binary orbital plane and the fundamental plane.
We will assume an optimal orientation of the detector, normal to the orbital plane, so that the orbit's inclination angle $\theta$ is zero. 
Then eq.~\eqref{eq:h_gen} will become:
\begin{equation}
\label{eq:h_plus}
h_{+} =-2\frac{M\eta}{R}
\bigg[ 
\left(-\dot r^2+r^2\dot \Phi^2+\frac{M}{r}\right) \cos 2 \Phi 
+ 2 r \dot r \dot \Phi \sin 2\Phi 
\bigg ], 
\end{equation}
\begin{equation}
\label{eq:h_cross}
h_{\times} =-2\frac{M\eta}{R}
\bigg[\left(-\dot r^2+r^2\dot \Phi^2+\frac{M}{r}\right) \sin 2\Phi 
- 2 r \dot r \dot \Phi \cos 2\Phi \bigg].
\end{equation}
The waveform strain is constructed from the plus and cross polarization modes as follows:
\begin{equation}
\label{eq:completeins}
h^{ins}(t) =  h_{+}(t) -i  h_{\times}(t).
\end{equation}
We mention that the variable $h$ (named strain or amplitude of the wave) is dimensionless, and represents the change in length divided by the length.
For our convenience, we chose to express equation \eqref{eq:completeins} in a more compact form, by transforming the sine and cosine terms into their exponential forms. 
The final result is: 
\begin{equation}
h^{\rm inspiral}(t) = A(t) e^{-i2\phi(t)} ~,
\text{and}~~
A = A_1+iA_2.
\label{eq:HA1A2}
\end{equation}
The amplitudes are given by
\begin{equation}
A_1 = -2\frac{M\eta}{R}\left(\dot{r}^2+r^2 \dot \Phi^2+\frac{M}{r}\right),~
\text{and}~~
A_2= -2\frac{M\eta}{R}\left(2 r \dot{r} \dot \Phi \right).
\label{eq:A1A2}
\end{equation}
\subsection{The Merger Model}
The merger starts beyond the innermost stable circular orbit (ISCO), which is defined as the last complete orbit before the binaries plunge and collide.  
The radius of this orbit is proved to be $r_{ISCO}=6 M$ in geometrical units (see Appendix~\ref{appx:Units}). 
The highly nonlinear merger phase is correctly modeled only by General Relativity, and numerical simulations of Einstein's equations are necessary to provide accurate results. 
However, because of the complexity and cost of numerical simulations, semi-analytical models were developed for the merger case, based on the results provided by numerical relativity. 
One of the most successful techniques is the \emph{Implicit Rotating Source} model, as is presented in~\cite{baker2008, east1212, huerta1609}, where the merger waveform is calculated by the analytical fit to numerical simulations.  
We build up the waveform strain for the merger starting with the equation: 
\begin{equation}
\label{m_model}
h^{\rm merger}(t) = A(t)\,e^{-i\Phi_{\rm gIRS}(t)}.\\
\end{equation}
We mention that we must use the formula given in~\cite{east1212} for the amplitude of the strain, because in~\cite{huerta1609} the power factor of ${1}{/2}$ seems to be missing:
\begin{equation}
\label{amp}
A(t)=\frac{A_0}{\omega(t)}\left[\frac{\big|\dot{\hat{f}}\big|}{1+\alpha\left(\hat{f}^2-\hat{f}^4\right)}\right]^{1/2},
\end{equation}
where:
\begin{equation}
\label{fhat}
\hat{f} = \frac{c}{2}\left(1+\frac{1}{\kappa}\right)^{1+\kappa}\left[1-\left(1+\frac{1}{\kappa}e^{-2t/b}\right)^{-\kappa}\right]\,.
\end{equation}
The angular orbital velocity is calculated with:
\begin{equation}
\label{ome}
\omega(t)=\omega_{\rm QNM}\left(1-\hat{f}\right).
\end{equation} 
where $\omega_{\rm QNM}$ is the fundamental, or least damped frequency of the quasi-normal modes (QNM) emitted by the final black hole as it settles into its spherical shape. 
We use for it the relation given in~\cite{huerta1609}: 
\begin{equation}
\label{omqnm}
\omega_{\rm QNM}= 1 - 0.63\left(1 - \hat{s}_{\rm fin}\right)^{0.3}
\end{equation}
where $\hat{s}_{\rm fin}$ is the spin of final black hole:
\begin{equation}
\label{fin_s}
\hat{s}_{\rm fin} = 2 \sqrt{3}\, \eta - \frac{390}{79} \eta^2 + \frac{2379}{287} \eta^3 - \frac{4621}{276} \eta^4\,.
\end{equation}
The phase is obtained by integrating the orbital angular velocity:
\begin{equation}
\label{int_phase}
\Phi_{\rm gIRS}(t) = \int_{t_0}^{t}\omega(t)\mathrm{d}t,
\end{equation}
The quantities: $\hat{f}$, $\dot{\hat{f}}={\mathrm{d} \hat{f}}/{\mathrm{d}t}$, and  $\hat{s}_{\rm fin}$ are obtained through an analytic fit to the numerical relativity results. 
The coefficients $\alpha$, $b$, $c$ and $\kappa$ are smooth function of the symmetric mass-ratio \(\eta\) and are given in Appendix~\ref{appx:Mcoeffs}. 
$A_0$ is a parameter which we can choose to be unity. 
This model applies to the merger of non-spinning compact binaries of different mass-ratios, and is called \emph{generic} IRS (gIRS) model.
We implement this simple model in \texttt{Mathematica} and use it to calculate the strain of the gravitational waves during the merger (see Appendix~\ref{appx:Procedures}).
\section{The Implementation of the Models}
\label{sec:Implementation}
\subsection{The Inspiral Gravitational Waveform}
Before starting the implementation of the models presented above, we need to determine the domain of the integration, ranging from an initial to a final value for the PN-parameter $x$. 
The lower boundary $x_0$ is dictated by the threshold value of the frequency when the signal enters the Advanced LIGO detection band. 
We consider this frequency as being determined by the cut-off frequency due to the Earth's seismic activity: $f^{low}_{GW}=10Hz$. 
It is worth mentioning that, because $x$ is a unitless physical parameter, we will calculate it using SI units.  
The orbital velocity corresponding to $f^{low}_{GW}$ is, from Kepler's third law, equal to
$v_0=\left( G {M} \omega^{low} \right)^{1/3}$, where $\omega^{low}=\pi f^{low}$, and ${M}$ is given in units of the solar mass $M_{\odot}$ (see Appendix~\ref{appx:Units}).
Note that the frequency of the gravitational waves is twice the orbital frequency $f_{GW} = 2 f_{orb}$, a known feature of quadrupole radiation.
This means that the gravitational wave signal goes through two maxima and two minima per one orbit of the binary motion~\cite{basicgw}.
With this expression for $v_0$, we can calculate the initial value for the PN parameter to be:
\begin{equation} 
x_0 = \left ( \frac{v_0}{c} \right)^2 = \left (\frac{G {M} \pi f_{GW}^{low}}{c^3}\right )^{2/3}. 
\end{equation}
The upper boundary is determined by the radius of the last stable orbit of the binary: $r_{ISCO} = 3 R_{Sch}=6 \frac{G {M}}{c^2}$. The velocity corresponding to this orbit in the Newtonian approach is: $v_{ISCO} = \sqrt {G{M}/r_{ISCO}}=c/\sqrt{6}$. With this value for the velocity, we obtain the upper limit for the PN parameter of $0^{th}$ order to be: $x^{0PN}_{ISCO}= 1/6$. We add a $2^{nd}$ order post-Newtonian correction, that introduces a dependence on the symmetric mass of the binary~\cite{blanchet13}, such that: 
 \begin{equation}
 \label{eq:xisco}
 x_{ISCO}^{2PN} = \frac{1}{6}\left( 1+\frac{7}{18}\eta\right).
 \end{equation} 
Next, using Kepler's third law we calculate the frequency of the gravitational wave at ISCO function the binary mass, by expressing the mass of the back hole in units of time (see Appendix~\ref{appx:Units}): 
\begin{equation}
f_{ISCO}= \sqrt{\frac{G{M}}{\pi^2 r_{ISCO}^3}}= \frac{c^3}{6^{3/2}G {M} \pi}.
\end{equation}
This shows that the frequency at the end of the inspiral scales inversely proportional to the mass of the binary, therefore the smaller the mass, the higher the frequency of the gravitational wave. 
After we determine the lower and upper bounds for the PN parameter $x$, we only need to choose a value for the symmetric mass ratio $\eta$, and then we can proceed with the integration of eq.~\eqref{eq:pre_hyb}. 

Let's start with a fiducial binary configuration of total mass $M=40$, and equal masses $m_1=m_2=20$ given in solar masses, corresponding to a symmetric mass ratio $\eta= 0.25$.   
Now we know everything and can solve numerically the differential equation~\eqref{eq:pre_hyb} using \texttt{Mathematica}. 
We set the lower boundary $x_{t=0}=x_0$. 
If we don't give an upper boundary for the time, we  run into a known issue in numerical analysis, where the step size becomes \emph{effectively zero}, and the equation becomes \emph{stiff}. 
Stiffness is a numerical property of differential equations, caused by a set of factors, such as the numerical method, initial conditions, or sudden changes in the solution due to singularities or sharp features in the solution for the differential equation. 
All this drives the step size to increasingly small values, until eventually it becomes effectively zero, which leads to unstable numerical results.  
We adopt the decision to go past the upper $x^{2PN}_{ISCO }$ boundary for $x$, and evolve to a final time very close to the time $t_{S}$ when the equation becomes stiff. 
Our choice is motivated by the fact that we included corrections up to 6PN order for $x$ in eq.~\eqref{eq:pre_hyb}, and supplemented with corrections up to 3PN for $r$ in eq.~\eqref{eq:r} so we can explore solutions beyond the last stable orbit.
To this extent we estimate the transition time from the stable binary black hole system to the coalescence into a single black hole, (also known as the \emph{time of flight}), which is the time necessary for the two black holes to fall from the last stable orbit (ISCO) to the light ring. 
This region is roughly located at twice the Schwarzchild radius: $r_{LR}=4M$ for a slowly rotating black hole.
While the event horizon is the invisible region around the black hole from which no light can escape, the light ring is made visible by the light forced to orbit around the black hole due to the lensing effect of the strong gravitational field. 
The distance from ISCO to the light ring is thus: $r_{ISCO} - r_{LR} = 2M$, and the corresponding time should be in geometrical units $t_{of}=2M$. 
We pick as the final time for the PN evolution the time $t_{F}=t_{S} - t_{of}$, and integrate numerically eq.~\eqref{eq:pre_hyb} from $t_0=0~\texttt{s}$ to $t_F$. 
Transformed in seconds by multiplication with $M_{\odot} (\texttt{s})$, the final time is $t_F=11.924~\texttt{s}$, and corresponds to a value $x_{final}=0.24585$, higher than $x^{2PN}_{ISCO}=0.18287$. 
We conclude that for a binary black hole of mass ${M} = 40 M_{\odot}$, the inspiral gravitational wave signal ideally stays in the detector range for a total time of nearly 12 seconds, and has a frequency range from $10~\texttt{Hz}$ to about $200~\texttt{Hz}$. 

\begin{figure}[th!]
    \centering
    \begin{subfigure}[b]{0.49\textwidth}
        \centering
         \includegraphics[width=\textwidth]{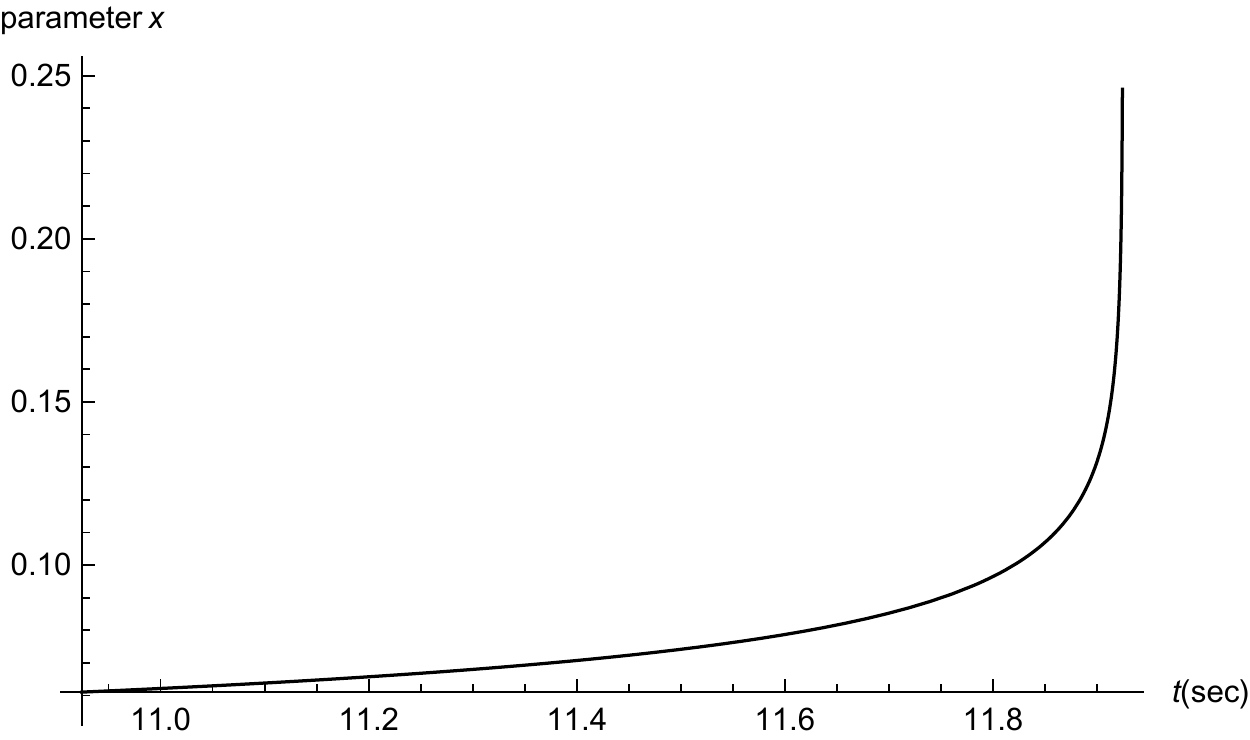}
    \end{subfigure}
    \begin{subfigure}[b]{0.49\textwidth}
        \centering
         \includegraphics[width=\textwidth]{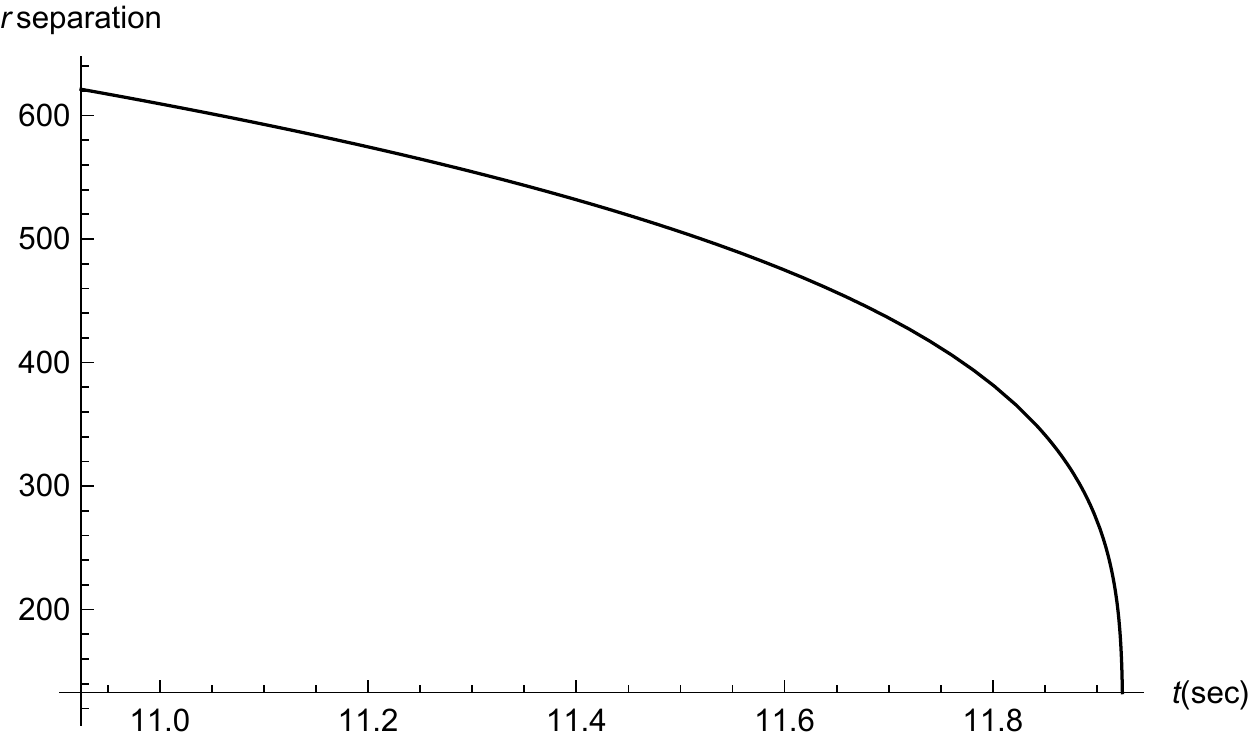}
    \end{subfigure}   
    \caption{The evolution in time of the PN parameter $x$ and separation $r$ for the inspiral of a black hole binary of total mass ${M}=40 M_{\odot}$}
    \label{fig:M40xrplot}
\end{figure}

After we obtain $x(t)$, we proceed to calculate the evolution of the phase with time, which is 
done by numerically integrating eq.~\eqref{eq:PhiEv}. 
Next, we need to determine the evolution of the distance between the orbiting black holes, known as the binary separation $r(t)$. 
For this, we use eq~\eqref{eq:r}, that gives $r(t)$ corrected up the $3^{rd}$ PN order in $x(t)$. 
We plot in Fig.~\ref{fig:M40xrplot} the evolution in time of the PN parameter $x(t)$ and of the separation $r(t)$, for the last second before the merger.
Before calculating the amplitude of the gravitational wave, let's assume a realistic value for the distance $R$ to the binary and take it to be the distance to Andromeda, the closest galaxy: $R= 2.4 \times 10^{19}~\texttt{km}$, or $2.5$ million light years.
The strain is a fractional quantity, therefore is unitless, and for consistency we must express the separation $r$ in seconds in eq.~\eqref{eq:h_plus}. 
We do this by multiplying it with the mass of the sun in seconds, as explained in Appendix~\ref{appx:Units}.
We also need to express the total mass $M$ in \texttt{km} if we want to take $R$ in \texttt{km}.
We calculate the plus and cross polarizations of the strain with eqs.~\eqref{eq:h_plus},~\eqref{eq:h_cross} and the amplitude with eq.~\eqref{eq:completeins}.
Our calculations show that the maximum amplitude of the gravitational wave strain is $A_{max}=|h^{inspiral}(t_F)|= 5.5 \times10^{-19}$, a shockingly small value. 

\begin{figure}[th!]
    \centering
    \begin{subfigure}[b]{0.47\textwidth}
        \centering
         \includegraphics[width=\textwidth]{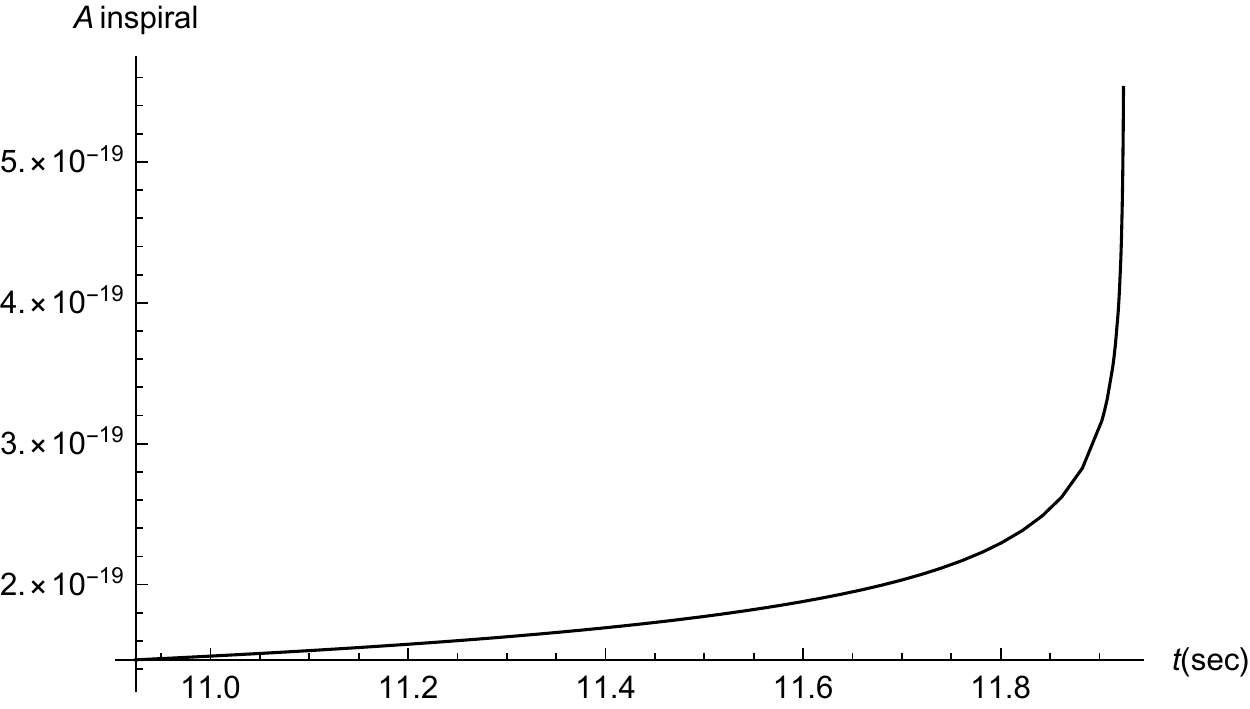}
    \end{subfigure}
    \begin{subfigure}[b]{0.52\textwidth}
        \centering
         \includegraphics[width=\textwidth]{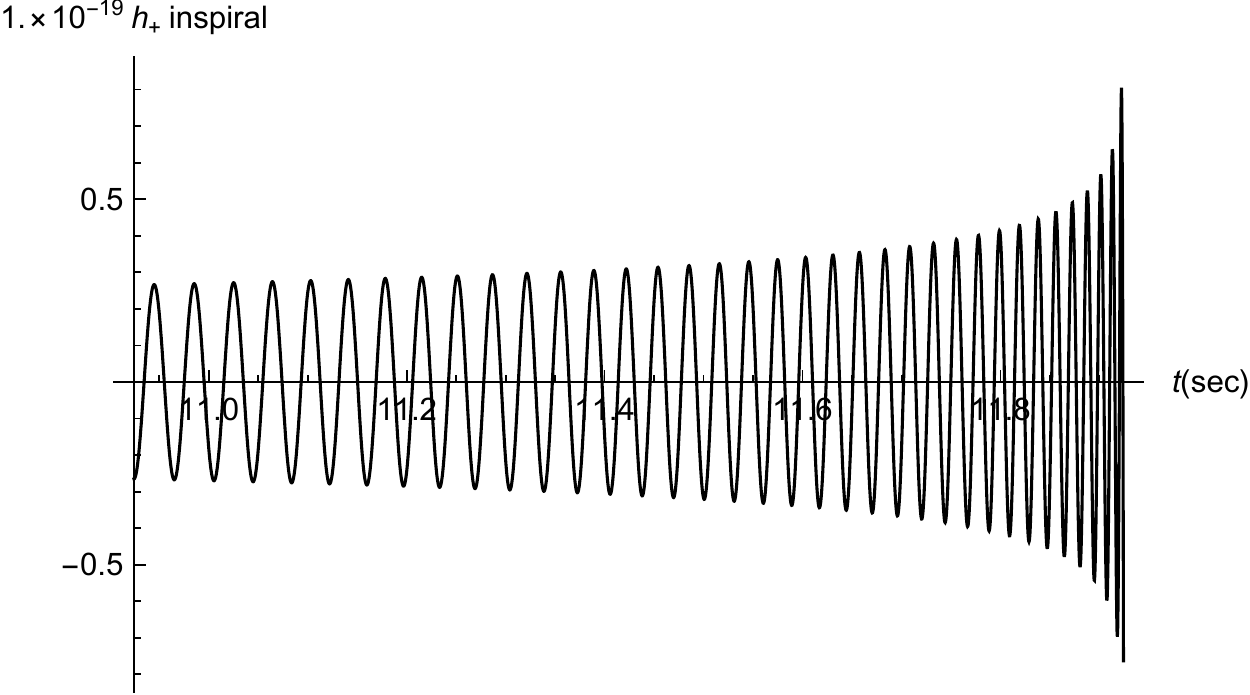}
    \end{subfigure}   
    \caption{The last second of the gravitational wave amplitude and $h_{+}$ component of the strain from the inspiral of a black hole binary of total mass ${M}=40 M_{\odot}$, at a distance $R= 2.4 \times 10^{19}\texttt{km}= 2.5$ million light years, or $778,000$ parsecs (Andromeda Galaxy).}
    \label{fig:M40Andromeda}
\end{figure}

This makes sense if we remember that the strain of the detected signal GW150914 was as small as $10^{-21}$, and the event was located at 1.3 billion light-years away! 
We plot in Fig.~\ref{fig:M40Andromeda} the amplitude and $h_{+}$ polarization mode of the gravitational wave for the last second before the merger, as it would be seen by an optimally oriented detector here on Earth.

\begin{figure}[th!]
    \centering
         \includegraphics[width=0.8\textwidth]{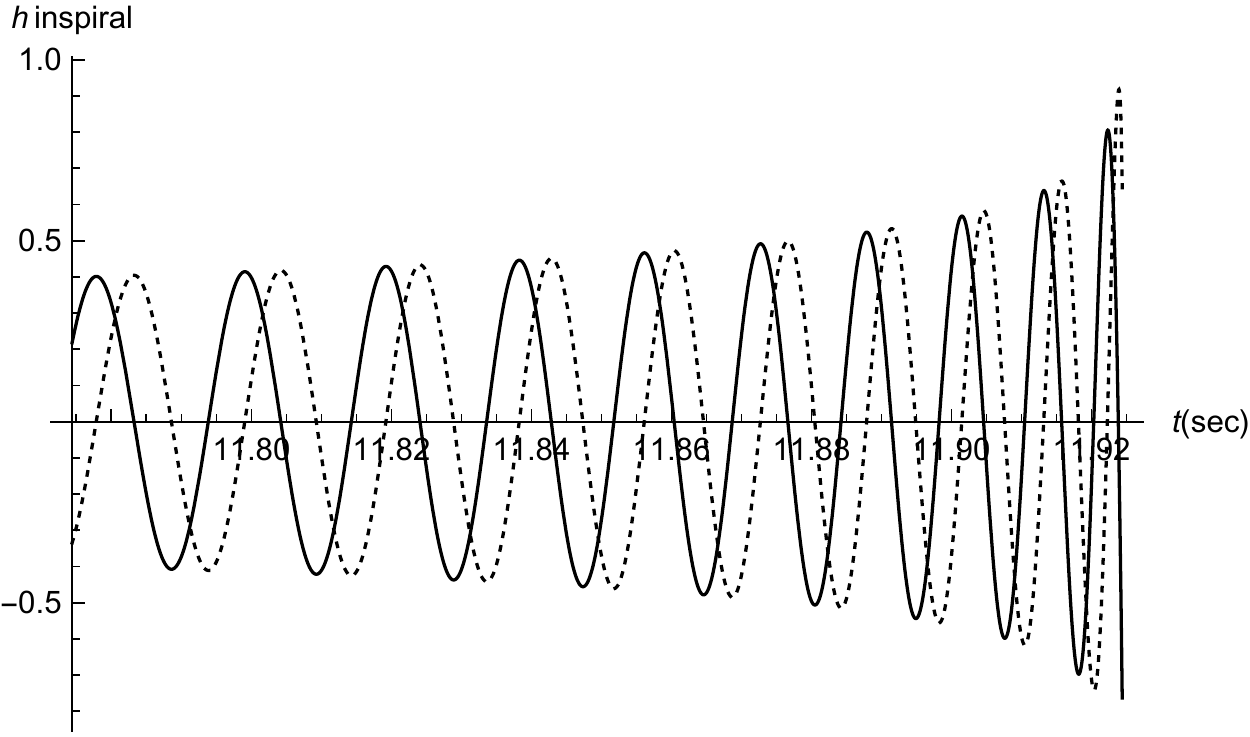}
    \caption{The evolution of the strain for the inspiral of an equal mass binary with total mass ${M}=40 M_{\odot}$. The solid plot represents the $h_{+}$, and the dashed plot is the $h_{\times}$ polarization.}
    \label{fig:M40h}
\end{figure}

The maximum amplitude of the strain will be $1$ and it's reached in the immediate vicinity of the binary black hole. 
We will rescale the strain to unity by dividing it with $A_{max}$, and plot it only the last $1/4$ of a second before the merger. 
We can see from Fig.~\ref{fig:M40h} that the two polarizations are in opposition of phase, and the $h_{\times}$ mode lags behind the $h_{+}$ mode.
Our next step is the calculation of the $h_{22}$ spherical harmonic component of the strain for the inspiral model. 
Spherical harmonic functions form a complete set of orthogonal functions defined on the surface of a sphere:
\begin{equation}
\label{eq:h22}
h_{22} =-4\frac{M\eta}{R}e^{-2i\Phi}\sqrt{\frac{\pi}{5}}\left( (r\dot \Phi +i \dot r)^2 +\frac{M}{r} \right). 
\end{equation} 
This is the dominant spherical harmonic mode in the gravitational wave signal. 
Indeed, we show in our \texttt{Mathematica} script that the difference between the strain calculated with eq.~\eqref{eq:h22} and with the eq.~\eqref{eq:HA1A2} is in the roundoff error, which is a proof to our calculations. 
\subsection{The Merger Gravitational Waveform}
\begin{figure}[th!]
    \centering
      \begin{subfigure}[b]{0.45\textwidth}
         \includegraphics[width=\textwidth]{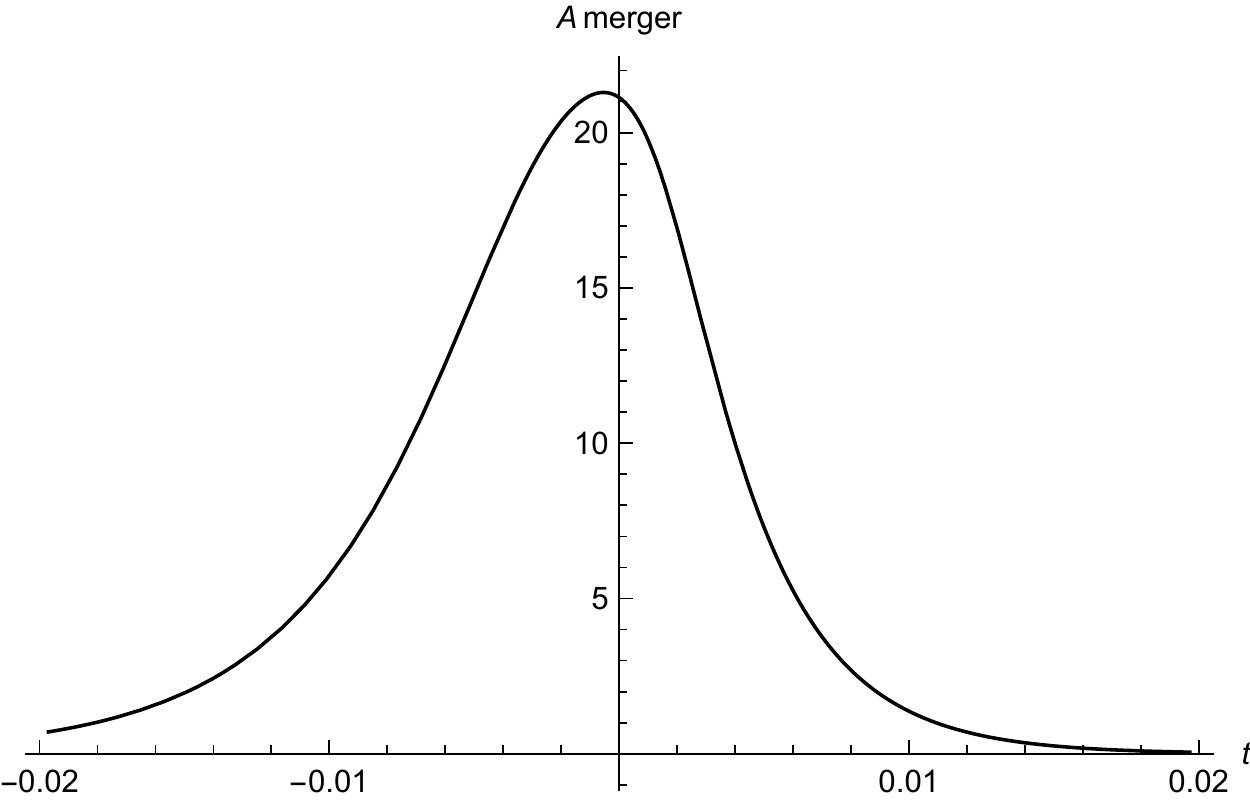}
     \end{subfigure}
        \begin{subfigure}[b]{0.54\textwidth}
         \includegraphics[width=\textwidth]{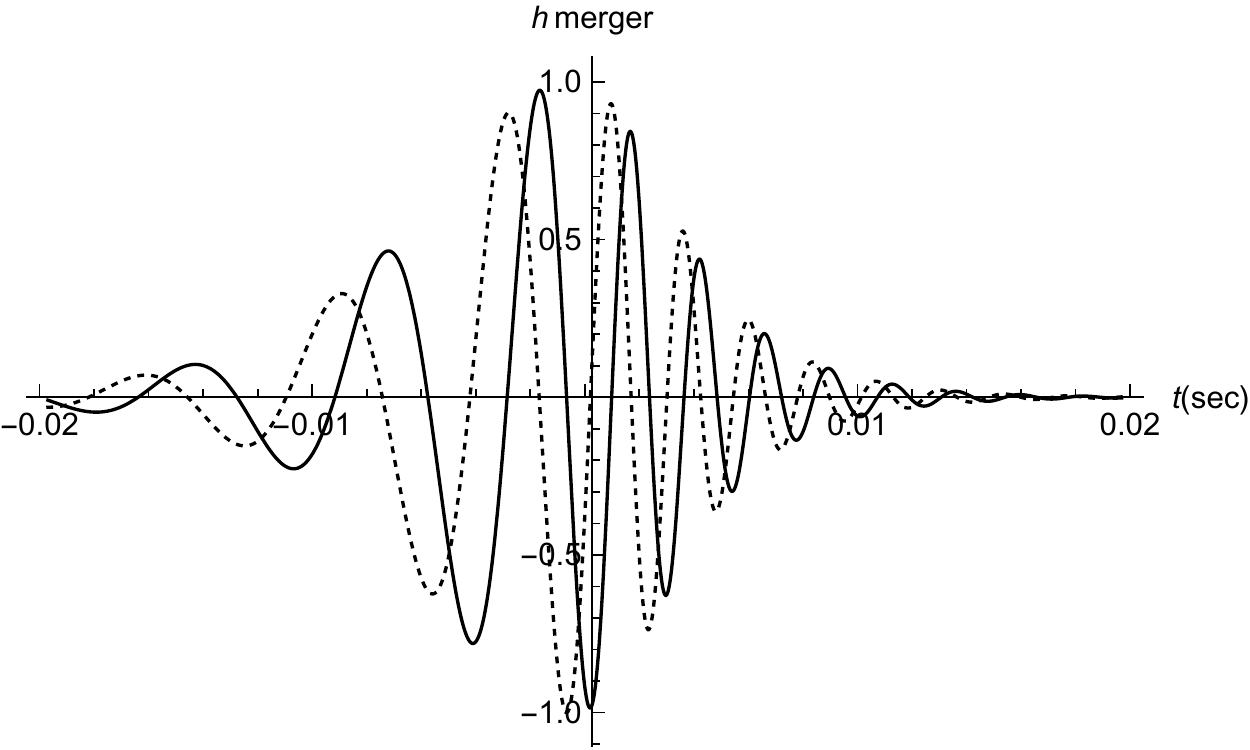}
        \end{subfigure}
    \caption{The amplitude and the rescaled strain for the merger of an equal mass binary of ${M}=40 M_{\odot}$. The solid plot represents the $h_{+}$, and the dashed plot is the $h_{\times}$ polarization.}
        \label{fig:M40Merger}
\end{figure}

The calculation of the gravitational wave strain for the merger proceeds in a straightforward way. 
We pick the same equal mass configuration for the binary, and start by calculating the angular frequency with eq.~\eqref{ome}, then we integrate it to obtain the phase $\Phi_{\rm gIRS}$.
Inspecting eq.~\eqref{fhat} we see that the time is in geometric units, and we replace it with $t \rightarrow \tfrac{t}{M M_{\odot}(\texttt{s})}$ in order to revert to time measured in seconds.
Next, we rescale the factor $A_0\rightarrow \tfrac{1}{M M_{\odot}(\texttt{s})}$ to render the strain unitless. 
With the amplitude given by eq.~\eqref{amp}, we compute the gravitational wave strain for the merger using eq.~\eqref{m_model}. 
We determine the maximum value for the amplitude of the merger strain, and rescale the amplitude to unity by dividing the strain with the maximum amplitude.
We plot in Fig.~\ref{fig:M40Merger} the evolution of the amplitude with time for a small time interval of $(-100M, +100M)$ around the origin, which expressed in seconds is around $(-0.02, 0.02)~\texttt{s}$. 
We see that the peak of the amplitude is not at $t_0=0$, but corresponds to a \emph {retarded time} $t_r=0.5388~\texttt{ms}$. 
We shift the merger strain with half that time: $t_r/2=0.269~\texttt {ms}$, bringing the highest maximum of the $h_{+}$ component to the time origin, and plot it in Fig~\ref{fig:M40Merger}. 
We see from Fig:~\ref{fig:M40Merger} that the $h_{+}$ mode lags behind the $h_{\times}$ polarization. 
\section{The Matching Technique}
\label{sec:Matching}
\subsection{Matching for a Template Waveform}
Next we will present a technique for constructing a complete gravitational waveform, by fitting together the strains for the inspiral and the merger, described in Sec.~\ref{sec:Implementation},  for the same binary configuration with total mass $M=40 M_{\odot}$.   
The gIRS waveform is tuned to numerical relativity results to model the dynamics of the merger, but we observe from Fig.~\ref{fig:M40Merger} that its amplitude diminished rapidly -- an indication that its accuracy deteriorates after only a few cycles, thus it has a very limited range of applicability. 
In developing our matching technique, we are relying on the accuracy of the inspiral evolution, which included corrections terms for the energy up to the $6^{th}$ order of the post-Newtonian approximation, and is evaluated up to the light ring.

\begin{figure}[th!]
    \centering
         \includegraphics[width=0.8\textwidth]{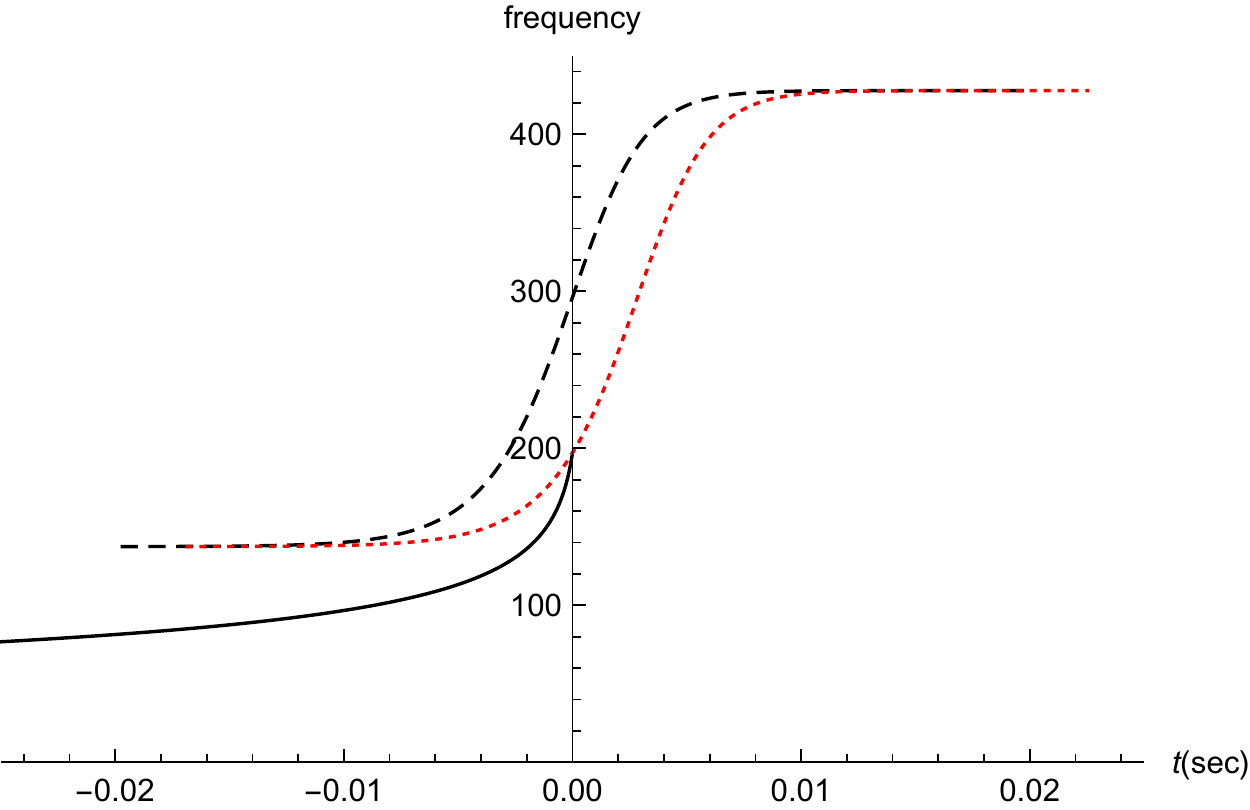}
    \caption{Comparison of the gravitational wave frequency at the end of inspiral and beginning of merger for an equal mass binary of ${M}=40 M_{\odot}$. 
    The solid line is frequency at the end of the inspiral, the long dashed line is the merger frequency, and the short dashed line (red) is the merger frequency shifted to overlap with the end inspiral frequency.}
        \label{fig:M40fGWshift}
\end{figure}

We start by determining the best matching interval by comparing the frequency evolution at the end of the inspiral and at the beginning of the merging phases. 
We calculate $f^{inspiral}_{GW} =\omega^{inspiral}/\pi $ and $f^{merger}_{GW} = \omega^{merger}/(2\pi)$. 
In fact, this relationship can be intuitively seen only comparing eq.~\eqref{eq:HA1A2} with eq.~\eqref{m_model} for the strain of the gravitational wave. 
In order to synchronize the two models at a time consistent with their common evolution, we first match them in frequency (see  Fig.~\ref{fig:M40fGWshift}). 
To do this, we shift the time axis of the inspiral frequency plot by $-t_F$ so that the end of the inspiral is at $t=0$, and then we shift the time axis of the gIRS frequency plot by a time parameter $\tau$, which is adjusted until the merger frequency at $t=0$ is approximately equal to the last frequency for the inspiral.
This makes the frequency plot continuous between the inspiral and merger phases. 
We see that the frequency of the merger phase increases abruptly, reaching more than twice the frequency at the end of the inspiral. 
This sudden increase in frequency during the coalescence is known as the \emph{chirp} of the gravitational wave. 
After this the binary enters into the ringdown phase, which ends when the final black hole is formed.
 Fig.~\ref{fig:M40fGWshift} shows the comparison between the shifted gravitational wave frequency at the end of the inspiral, the merger frequency, and the translated merger frequency with time $\tau$, until the overlap with the frequency of the inspiral is reached. 
This is a straightforward technique and can be easily used in lab settings, or in hands-on demonstrations on gravitational waves. 
We obtain the frequency overlap for a time shift $\tau = 2.84~\texttt{ms}$, and apply this time shift to both polarization modes of the merger.

\begin{figure}[th!]
    \centering
    \begin{subfigure}[b]{0.75\textwidth}
         \includegraphics[width=\textwidth]{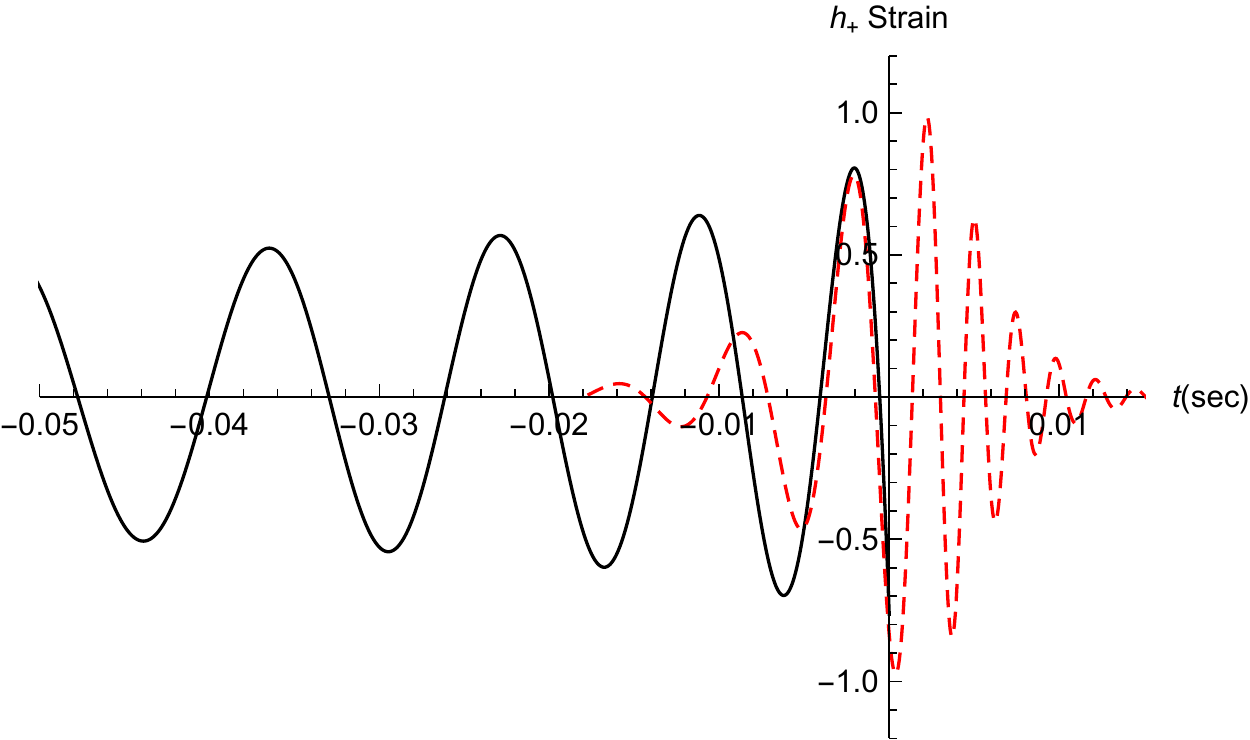}
    \end{subfigure}
    \begin{subfigure}[b]{0.75\textwidth}
         \includegraphics[width=\textwidth]{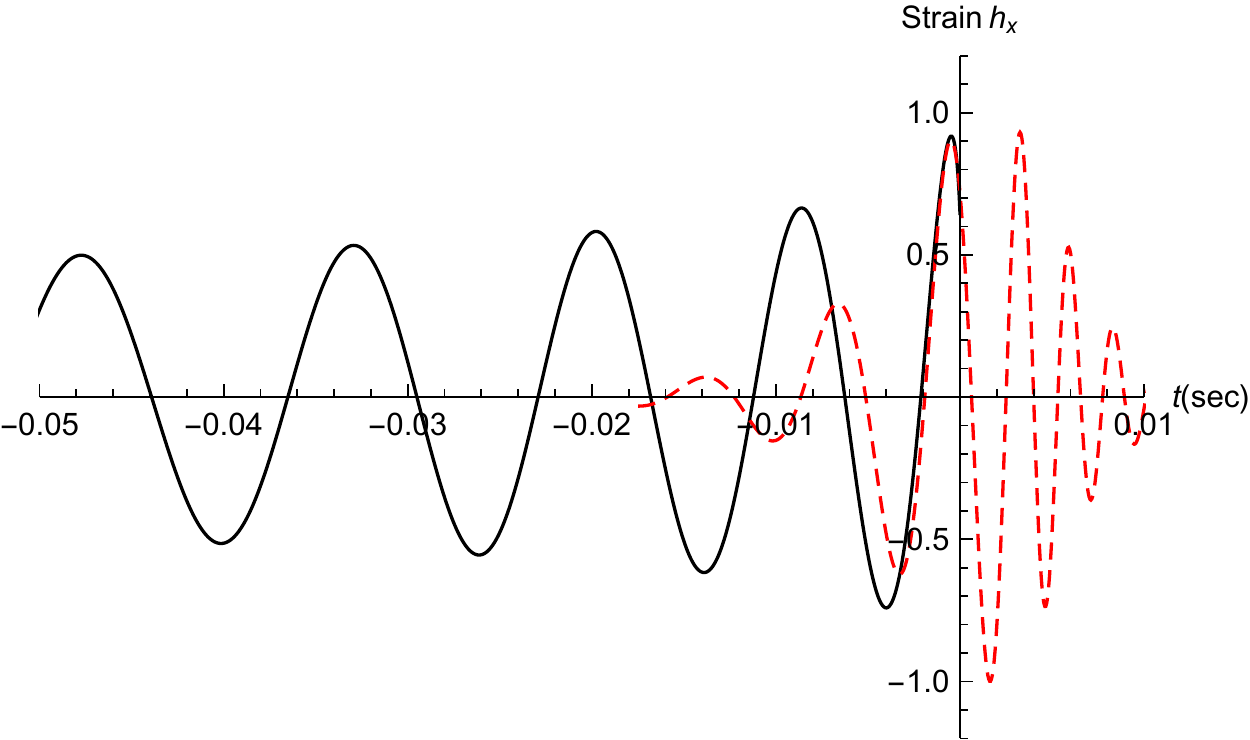}
     \end{subfigure}
            \caption{The overlap of the strain for an equal mass binary of total mass ${M}=40 M_{\odot}$. The solid black plot represents the inspiral strain, and the dashed red plot is the merger strain.}
        \label{fig:M40Match}
\end{figure}

Next we proceed to match the inspiral and merger strains. 
We analyze first the $h_{+}$ polarization and observe that we need to account for the phase difference between the inspiral and the merger, because the inspiral $h_{+}$ mode leads, while the merger $h_{+}$ lags. 
We find that a strain parameter $\Phi_0=\pi$ will bring the  inspiral and merger $h_{+}$ strain in phase. 
We obtain a clear overlap at the last maximum of the inspiral strain, then adjust the time axis in increments of the retarded time $t_r/2$, which is about a tenth the time shift $\tau$, until the peak of the inspiral overlaps with the peak of the merger waveform. 
Remarkably, with only a time shift of $\Delta t = 3/2 t_r$ we obtain a very good fit and we do not have to rescale the amplitude of the strain at the overlapping point. 
The $h_{\times}$ polarization is not affected by a phase difference, and an excellent overlap is obtained when we adjust the time axis with only $\Delta t = t_r$. 
We see from Fig.~\ref{fig:M40Match} that the matching interval is optimal, because we can pick other points in the vicinity of the peak and obtain the same high overlap between the inspiral and the merger amplitudes. 
\subsection{Comparison with the GW150914 Waveform}
We test our implementation by calculating the inspiral and merger strain for the binary configuration GW150914, and comparing our results with the gravitational-wave strain template for this event, released by the Gravitational Wave Open Science Center ~\cite{LOSC}.
This procedure will test both our implementation of the post-Newtonian and gIRS models and our overlapping technique, confirming its viability.  
The mass parameters for this binary configuration are $m_1 = 36.2 M_{\odot}$ and $m_2 = 29.1 M_{\odot}$, with a symmetric mass ratio $\eta=0.247$, close to the one of an equal mass binary.
The total mass of the remnant black hole is ${M}=62.3 M_{\odot}$, which gives a Schwarzchild radius of only $1.8\times 10^2~\texttt{km}$, while the distance to the event is estimated to be about $1.2\times 10^{22}~\texttt{km}$.
Using the technique described in Sec.~\ref{sec:Implementation}, with the value for the cutoff detector frequency $f^{low}_{GW}=10~\texttt{Hz}$, we obtain the strain of the gravitational wave for the inspiral and merger phases of the binary evolution.
The final time for the inspiral strain $t_F=5.126~\texttt{s}$ and its maximum amplitude is $A_{max}=1.78\times 10^{-21}$. 

\begin{figure}[th!]
    \centering
    \begin{subfigure}[b]{0.75\textwidth}
         \includegraphics[width=\textwidth]{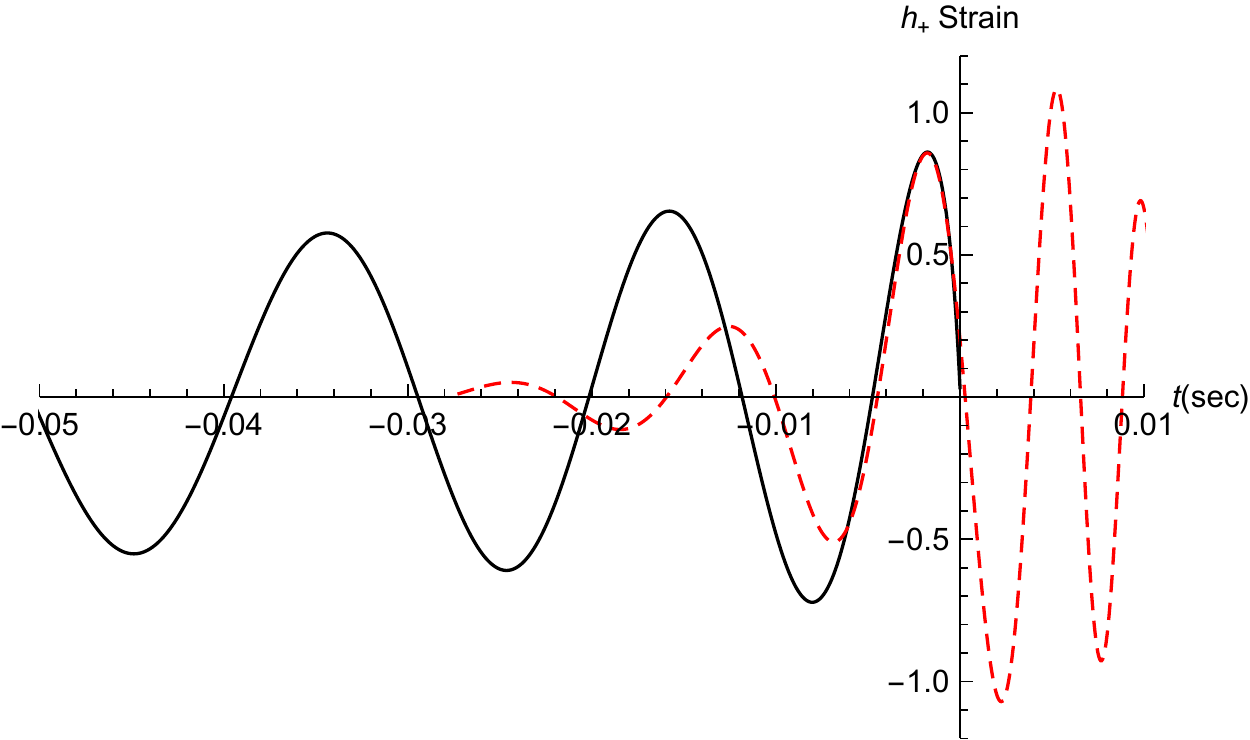}
    \end{subfigure}
    \begin{subfigure}[b]{0.75\textwidth}
         \includegraphics[width=\textwidth]{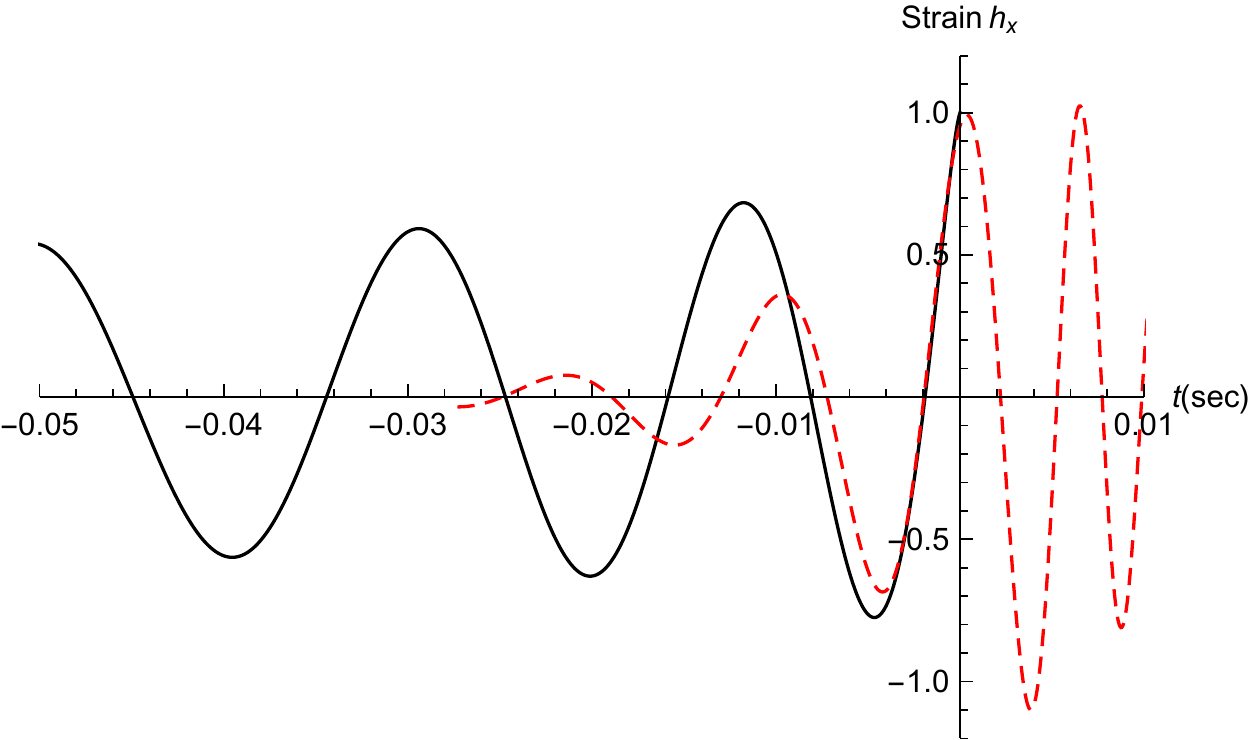}
    \end{subfigure}
    \caption{The overlap of the strain for the binary of total mass ${M}=65.3 M_{\odot}$. The solid black plot represents the inspiral strain, and the dashed red plot is the merger strain.}
        \label{fig:M65Match} 
\end{figure}

The merger strain is calculated within a time interval of $(-0.032, 0.032)~\texttt{s}$ and the retarded time for the peak in amplitude is $t_r=0.8747~\texttt{ms}$.  
The frequency overlap is obtained for a time shift $\tau = 4.5~\texttt{ms}$, which is used to shift the merger strain before matching. 
Fig.~\ref{fig:M65Match} shows the complete waveforms for the $h_{+}$ and $h_{\times}$ polarizations, obtained with $\Delta t = t_r/2$ for both modes. 
We increase the amplitude of the merger waveform with 10\%, and correct the phase of the $h_{+}$ mode with $\Phi_0=\pi$, to obtain a very good overlap between the inspiral and merger. 

\begin{figure}[th!]
    \centering
         \includegraphics[width=0.9\textwidth]{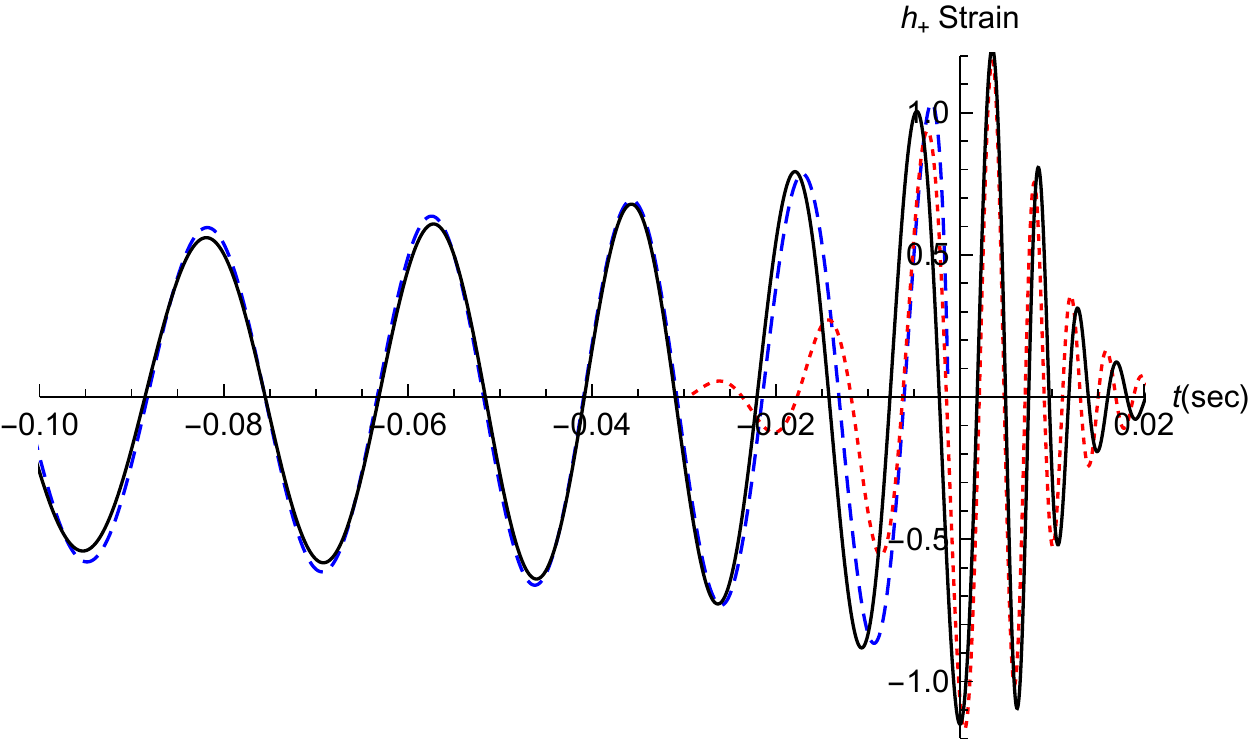}
    \caption{Comparison between the GW150914 template and our complete model for the $h_{+}$ polarization model.
     The solid black line is the GW150914 template, the long dash blue line is the inspiral, and the small dash red line is the merger model. 
    Our model was shifted in time by $\Delta t = 1.312~\texttt{ms}$ and its amplitude was increased by 20\%.} 
        \label{fig:G150914Match}
\end{figure}

Lastly, we read in \texttt{Mathematica} the data for the open source GW150914 template, then overlap it with our calculated strain for the ${M}=65.3 M_{\odot}$ binary. 
Fig.~\ref{fig:G150914Match} shows an excellent match between our post-Newtonian waveform and the GW150914 template, for several peaks, with a time adjustment of only $\Delta t = 3/2 t_r$ and an increase in amplitude of 20\%. 
Our model loses accuracy at the last peak, and this is expected, because we pushed our calculation beyond the limit of its applicability. 
The gIRS strain for the merger is adjusted with $\Delta t = 3/2 t_r$, which gives a good overlap with the GW150914 template at peak amplitude.

\section{Conclusions}
\label{sec:Conclusions}
Our objective of developing a streamlined matching method for the analytical calculation of the complete gravitational waveform, while keeping it simple enough to be accessible to undergraduate physics students, was accomplished. 
We implemented two analytical algorithms for calculating gravitational wave templates during the inspiral and merger of compact binary systems, and we built a cohesive method of combining them into a complete waveform. 
We bypass the complicated Einstein's equations by using the post-Newtonian (PN) theory to model the inspiral phase and the Implicit Rotating Source (IRS) for the merger phase of the binary evolution. 
After building the inspiral and merger waveforms, we devised our matching method and validated it by comparing our results with the waveform template for GW150914, the first detection of gravitational waves. 
This is a rich and timely topic, and our approach, accessible to undergraduate students, can easily be implemented in a special topics course or research project for a junior or senior physics students.
We provide the \texttt{Mathematica} scripts and explain in Appendix~\ref{appx:Procedures} the start-up procedure to be followed by beginners in this field in order to generate a complete waveform.  
There are several future projects that can be developed based on this report, among which are building complete waveforms for all the detected signals, extending the inspiral mode to include non-zero eccentricity, testing,  improving and optimizing the matching technique, employing and testing a more realistic analytic model for the merger, adding a ringdown model, etc.
\vfill
\appendix   
\section{Geometrical Units}
\label{appx:Units}
Throughout this paper we are working in \emph{Geometrical Units} (GU), in which calculations are simplified, because we don't have to deal with physical constants such as the universal gravitational constant (\textit{G}) or the speed of light (\textit{c}), because they are set to unity ($\textit{G}=\textit{c}=1$). 
Let's first set the speed of light to unity. Then we can measure time in units of distance:
\begin{equation}
 \textit{c}=1 = 2.998\times10^8~\texttt{m/s}\rightarrow
  1~\texttt{s} = 2.998\times10^8~\texttt{m},
 ~~ 1~\texttt{m} = 3.336\times10^{-9}~\texttt{s}. 
  \label{eq:A1}
\end{equation}
Now by setting Newton's gravitational constant to unity, and taking the unit for time as measured in meters, we can measure mass in units of distance as well:
\begin{equation}
 \textit{G} = 1 = 6.673\times10^{-11}~\texttt{m}^3/~\texttt{kg}\cdot~\texttt{s}^2\rightarrow
 1~\texttt{kg} = 0.742\times10^{-27}~\texttt{m} = 0.742\times10^{-30}~\texttt{km}. 
 \label{eq:A2}
\end{equation}
Therefore, we measure both mass and time in $\texttt{meters}$, which is a distance, or a \emph{geometrical} unit.
In order to establish a straightforward correspondence between theoretical geometrical units and observations, we express all the relevant quantities in units of solar masses and use this quantity when converting back and forth between those systems of units.   
To this purpose, let's calculate the mass of the Sun, $M_{\odot}$, in $\texttt{km}$ and in $\texttt{s}$, by using the tricks we explained above in eq.~\eqref{eq:A1} and eq.~\eqref{eq:A2} to transform $\texttt{kg}$ into $\texttt{km}$ and into $\texttt{s}$. 
Thus $M_{\odot}$ can be used as a universal unit, that measures mass, distance, and even time. 
\begin{equation}
 M_{\odot}=1.989\times10^{30}~\texttt{kg} = 1.476~\texttt{km} = 4.923 \times10^{-6}~\texttt{s}.
 \label{eq:A3}
 \end{equation}
Let's explain how can we use $M_{\odot}$ as the unit for mass, distance and time, with this conversion, by giving a few examples. 
The mass of a black hole measured in solar masses is simply $m {M_{\odot}}$, where $m$ is a dimensionless multiplication number, and ${M_{\odot}}$ is the mass of the Sun in \texttt{kg}. 
As another example, let's recall that the radius of a black hole for which the escape speed is equal to the speed of light, called the Schwarzchild radius, is $R_{Sch} = 2\frac{G}{c^2} M$. 
The Schwarzchild radius becomes simply $R_{Sch} = 2 M$ in geometrical units. 
For the mass of the Sun, this radius is $R_{Sch,\odot}=2.95~\texttt{km}$. 
Now, by only expressing the black hole mass in terms of the solar mass ${M_{\odot}}$ written as unit of distance, we get back to the SI units: $R_{Sch} = 2 M M_{\odot} (\texttt{km})= 2.95 M~\texttt{km} $. 
\section{Post-Newtonian Coefficients}
\label{appx:PNcoeffs}
We give below the coefficients used in eq.~\eqref{eq:pre_hyb} for the calculation of the post-Newtonian variable $x$. 
\begin{align*}
a_4 &=170.799 - 742.551 \eta + 370.173 \eta^2 - 43.4703 \eta^3 - 
 0.0249486 \eta^4 + (14.143 - 150.692 \eta) \log (x(t)) \\
a_{9/2} &=  1047.25 - 2280.56 \eta + 923.756 \eta^2 + 22.7462 \eta^3 - 
 102.446 \log (x(t)) \\
a_5 &= 714.739 - 1936.48 \eta + 3058.95 \eta^2 - 514.288 \eta^3 + 
 29.5523 \eta^4 - 0.185941 \eta^5 \\
 &+ (-3.00846 + 1019.71 \eta + 1146.13\eta^2) \log (x(t)) \\
a_{11/2} &= 3622.99 - 11498.7  \eta + 12973.5  \eta^2 - 1623.  \eta^3 + 
 25.5499  \eta^4 + (83.1435 - 1893.65  \eta) \log (x(t)) \\
a_6 &= 11583.1 - 45878.3 \eta + 33371.8 \eta^2 - 7650.04 \eta^3 + 
 648.748 \eta^4 - 14.5589 \eta^5 - 0.0925075 \eta^6 \\
 &+ (-1155.61 + 7001.79 \eta - 2135.6 \eta^2 - 
    2411.92 \eta^3) \log (x(t)) + 33.2307 \log (x(t)^2)
\end{align*}
Below are coefficients used in eq.~\eqref{eq:r} for the calculation of the separation $r(t)$ as an expansion in the post-Newtonian variable $x$. 
\begin{align*}
r0PN &= 1 \\
r1PN & = -1 + 0.333333 \eta \\
r2PN &= 4.75 \eta + 0.111111 \eta^2 \\
r3PN &= -7.51822 \eta - 3.08333 \eta^2 + 0.0246914 \eta^3
\end{align*}
\section{gIRS Coefficients}
\label{appx:Mcoeffs}
Here are the coefficients used in the calculation of the merger waveform.
\begin{align*}
Q\left(\hat{s}_{\rm fin}\right) & =  \frac{2}{\left(1-\hat{s}_{\rm fin}\right)^{0.45}} \\
\alpha\left(\eta\right)&=\frac{1}{Q^2\left(\hat{s}_{\rm fin}\right)}\left(\frac{16313}{562} + \frac{21345}{124}\eta\right) \\
b\left(\eta\right)&=\frac{16014}{979} - \frac{29132}{1343}\eta^2 \\
c\left(\eta\right)&=\frac{ 206}{903} + \frac{180}{1141}\sqrt{\eta} + \frac{424}{1205}\frac{\eta^2}{\log\left(\eta\right)} \\
\kappa\left(\eta\right)&=\frac{713}{1056}-\frac{23}{193}\eta
\end{align*}

\section{Procedures}
\label{appx:Procedures}
We lay out below the step-by-step procedure to be followed by beginners in this field, including students with little or no knowledge of gravitational waves or \texttt{Mathematica}, in order to generate complete waveforms. 
One word of caution: this model is tailored to work best for binary configurations of comparable mass ratio.  
Experiment with it and please do not hesitate to send an email to the first author if you run into a problem.
\begin{enumerate}[noitemsep,topsep=0pt]
\item \underline{Setting up}: Install the \texttt{Mathematica}~\cite{MATH} software, then go to \url{https://github.com/mbabiuc/MathScripts}, click on the \fbox{Clone or Download} button on the right, and choose the \fbox{Download ZIP} option.  
\item \underline{Mass parameters}: Open the \texttt{Match40.nb} script on your computer and save it under a different name. Then under the section \emph{Setting up the Mass Parameters} change the mass parameters \texttt{m1} and \texttt{m2} with values of your choice, and run the script (click on \fbox{Evaluation} and chose \fbox{Evaluate Notebook}).
\item \underline{Final integration time}: Scroll down to the section titled \emph{Setting up the final integration time} text line, and note the error given by the function \texttt{NDSolve: At t == ..., step size is effectively zero; singularity or stiff system suspected}.
Copy and paste the numerical value for that time to the \texttt{tS} variable defined immediately below the \texttt{NDSolve} and run the script again.
\item\underline{Inspiral Waveform}: Scroll to the section \emph{Calculation of the Inspiral Waveform} and if you want, change the variable \texttt{R} to a realistic value for the distance from the detector the to source. Now run only that portion of the script again (press \fbox{Shift}+\fbox{Enter}) until the section \emph{Calculations for Merger-Ringdown Waveform}. 
Congratulations, you just generated your first gravitational wave model for the inspiral! 
If you want to save the plot, you will type in the notebook, just below the plot, the command \texttt{Export["/Path\_of\_File/Name\_of\_File", \%, "PDF"]} and run it.
\item\underline{Merger Waveform}: This is generated without any intervention.
\item\underline{Matching in frequency}: In the \texttt{Manipulate} plot, click the $+$ button, then play the graph until the plot goes through the origin of the axes. 
The shift is done in increments of the retarded time $t_r$. 
Divide the value obtained for $\tau$ by $t_r$, to obtain the factor $\chi$, and check if the frequency of the merger at that time $f_M$ is nearly equal to the last frequency of the inspiral $f_F$. 
If not, tweak the factor $\chi$ to obtain the best concordance.
\item\underline{Matching in amplitude}: Lastly, the matching in amplitude should follow straight from the matching in frequency. 
It might require a slight adjustment of the time axis for the merger model in increments of $t_r$. This is done by tweaking the $\epsilon_{+}$ and $\epsilon_{\times}$coefficients.
\end{enumerate}
We hope that this procedure will be easy to follow and rewarding, and will be useful in bootstrapping future projects in gravitational waves with undergraduates, increasing the involvement of the physics students and faculty in this new and exciting field. 

\vfill
\begin{acknowledgments}

This work was supported by the department of Physics at Marshall University and the NSF EPSCoR Grant OIA-1458952 to the state of West Virginia ``Waves of the Future".

\end{acknowledgments}

\end{document}